\documentclass[11pt,twoside]{article}
\usepackage{./asp2014}

\aspSuppressVolSlug
\resetcounters

\begin{document}

\title{The Distribution of Supernovae Relative to Spiral Arms of Host Disc Galaxies}
\author{L.~S.~Aramyan,$^1$ A.~A.~Hakobyan,$^1$ A.~R.~Petrosian,$^1$ V.~de~Lapparent,$^{2}$
        E.~Bertin,$^{2}$ G.~A.~Mamon,$^2$ D.~Kunth,$^2$
        T.~A.~Nazaryan,$^{1}$ V.~Adibekyan,$^3$ and M.~Turatto$^4$
\affil{$^1$Byurakan Astrophysical Observatory, Byurakan, Armenia;
       \email{aramyan@bao.sci.am}\\
       \email{hakobyan@bao.sci.am}\\
       $^2$Institut d'Astrophysique de Paris, Paris, France;\\
       $^3$Instituto de Astrof\'{i}sica e Ci\^{e}ncia do Espa\c{c}o, Porto, Portugal;\\
       $^4$Osservatorio Astronomico di Padova, Padova, Italy}}

\paperauthor{L.~S.~Aramyan}{}{}{Byurakan Astrophysical Observatory}{}{Byurakan}{}{}{Armenia}
\paperauthor{A.~A.~Hakobyan}{hakobyan@bao.sci.am}{}{Byurakan Astrophysical Observatory}{}{Byurakan}{}{}{Armenia}
\paperauthor{A.~R.~Petrosian}{}{}{Byurakan Astrophysical Observatory}{}{Byurakan}{}{}{Armenia}
\paperauthor{V.~de~Lapparent}{}{}{Institut d'Astrophysique de Paris}{}{Paris}{}{}{France}
\paperauthor{E.~Bertin}{}{}{Institut d'Astrophysique de Paris}{}{Paris}{}{}{France}
\paperauthor{G.~A.~Mamon}{}{}{Institut d'Astrophysique de Paris}{}{Paris}{}{}{France}
\paperauthor{D.~Kunth}{}{}{Institut d'Astrophysique de Paris}{}{Paris}{}{}{France}
\paperauthor{T.~A.~Nazaryan}{}{}{Byurakan Astrophysical Observatory}{}{Byurakan}{}{}{Armenia}
\paperauthor{V.~Adibekyan}{}{}{Instituto de Astrof\'{i}sica e Ci\^{e}ncia do Espa\c{c}o}{}{Porto}{}{}{Portugal}
\paperauthor{M.~Turatto}{}{}{Osservatorio Astronomico di Padova}{}{Padova}{}{}{Italy}

\begin{abstract}
  Using a sample of 215 supernovae (SNe), we analyse their
	positions relative to the spiral arms of their host galaxies,
	distinguishing grand-design (GD) spirals from non-GD (NGD) galaxies.
	Our results suggest that shocks in spiral arms of GD galaxies
	trigger star formation in the leading edges of arms
	affecting the distributions of core-collapse (CC) SNe (known to have
	short-lived progenitors). The closer locations of SNe~Ibc vs. SNe~II
  relative to the leading edges of the arms supports the belief
	that SNe~Ibc have more massive progenitors. SNe~Ia having less massive
	and older progenitors, show symmetric distribution
	with respect to the peaks of spiral arms.
\end{abstract}

\section{Introduction}

It is well known that star forming regions in spiral discs are generally concentrated
in spiral arms \citep[e.g.][]{2002MNRAS.337.1113S}.
There are a variety of known structures of spiral galaxies,
with different numbers and shapes of
their arms \citep[for recent review see][]{2013pss6.book....1B}.
According to their spiral features, spiral galaxies are divided into two broad categories:
1) GD spirals with typically two arms; and
2) NGD spirals with flocullent and/or short arms.
Spiral arms in GD galaxies are thought to be density waves that are usually attributed to
stellar bars \citep[e.g.][]{1976ApJ...209...53S,1989ApJ...342..677E,2013JKAS...46..141A,2013MNRAS.432.2878R},
or to the tidal field of a nearby neighbor
(e.g. \citealt{1972ApJ...178..623T,1979ApJ...233..539K,2011MNRAS.414..538K,2013MNRAS.429.1051C}).
In contrast to GD spiral galaxies, NGD galaxies
are likely formed from gravitational instabilities,
or are sheared star formation regions
(e.g. \citealt{1982FCPh....7..241S}; \citealt*{2003ApJ...590..271E}).

The distribution of stellar ages in spiral arms
have been studied in GD galaxies.
Investigating the dynamics of spiral galaxies,
\citet{1969ApJ...158..123R} proposed that the piled up gas
in a spiral arm experiences a strong shock
that triggers star formation. Then, many studies with contradicting results
came out that observationally support or argue the proposed theory.

By this contribution, the fourth study of a series
\citep[][]{2012A&A...544A..81H,2014MNRAS.444.2428H,2016MNRAS.456.2848H,2016MNRAS.459.3130A},
we present our recent investigation of the distribution of different types of SNe
relative to spiral arms taking into account the intrinsic properties of arms in order to
find links between the distributions of the various SN types and arm's stellar populations.
Moreover, considering possible differences of the distributions of
various stellar populations in both types of spirals \citep[e.g.][]{2010MNRAS.409..396D},
we investigate the distribution of SNe in these subsamples of host galaxies.

\section{The sample of SNe and their hosts}

The sample of this study is drawn from the catalog
of \citet{2012A&A...544A..81H}, which contains 3876 SNe
from the area covered by the Sloan Digital Sky Survey (SDSS)
Data Release 8. Since, the analysis of the distribution
of SNe relative to spiral arms requires a well-defined sample
and high angular resolution images of the SNe hosts, we
applied some restrictions and came out with a final sample consisting 215 SNe in 187 host galaxies.
The spiral arm classes of all 187 host galaxies were determined
visually from the $g$-band SDSS images, following the spiral arm 
classification of \citet{1987ApJ...314....3E}.
Then the galaxies were assigned as GD (classes 9 and 12) or
non-GD (NGD, all classes except 9 and 12). 

Using the version~2.18.4 of {\sc SExtractor}
software \citep{1996A&AS..117..393B},
we carried out a procedure to isolate the spiral structure of the galaxies.
We first fitted all $g$-band SDSS images of the host galaxies 
in the sample with bulge+disc models ($r^{1/4}$ bulge
and exponential disc profiles are used for all galaxies).
The modeled bulge+disc was then subtracted from each original image.
Then, the closest segment of spiral arm to SN position is fixed along the radial line passing
through the galaxy nucleus and SN location and the radial light profile is obtained.
Throughout whole study we used $d_1$ and $d_2$ distances, which indicate the position
of SN relative to the edges and peak of spiral arm, respectively.
For more details the reader is referred to \citet[][]{2016MNRAS.459.3130A}.

\section{Results}

It is believed that the spiral arms of NGD galaxies corotate
with the discs and do not show signs of shocks
in their leading edges \citep[see review by][]{2014PASA...31...35D}.
Therefore, one does not expect young stars to be
concentrated towards one of the edges of their arms.
This scenario also predicts
the absence of radial trends for the distributions of SNe~Ibc and II
inside the spiral arms.
In NGD galaxies, we found that
CC SNe are more concentrated towards the peaks
of spiral arms than SNe~Ia.
Moreover, despite small number statistics, in NGD galaxies
we found different concentration levels for SNe~Ibc and II.
In particular the mean absolute $d_2$ distance
of SNe~Ibc is $0.27 \pm 0.08$ ($N = 13$)
and for SNe~II is $0.49 \pm 0.07$ ($N = 39$).
An Anderson-Darling (AD) test shows that the difference between the distributions
of absolute distances of SNe~Ibc and II from the peaks of spiral arms
is barely significant ($P_{\rm AD} = 0.074$).
Hence, in NGD galaxies the shortest mean distance
to the peak of spiral arm is for SNe~Ibc.
In addition, the distribution of any SN type inside the spiral arms
in NGD galaxies does not show any significant radial trend.

Assuming that 1) the \textit{g}-band profiles of spiral arms
represent the distribution of young stars, and 2) the
peaks of spiral arms of NGD galaxies are the most suitable sites of the
star formation, we propose that when the concentration
of a given type of SNe towards the arm peak is higher,
their progenitors are younger (more massive in the context of single-star evolution).
Thus, a mass sequence Ia--II--Ibc for the SN progenitors is expected,
in agreement with those from the literature obtained by
the association of various SN types with the
$\rm H\alpha$ emission of the host galaxy
\citep[e.g.][]{2006A&A...453...57J,2008MNRAS.390.1527A,2012MNRAS.424.1372A}.

The distribution of SNe inside the spiral arms
of GD galaxies is quiet different.
In particular, in GD galaxies we found a significant shift between
the distributions of distances of CC and Ia SNe from the peaks of spiral arms.
This is probably the reflection of the offset between \textit{B}-
and \textit{I}-band light profiles of spiral arms with strong density waves
reported by \citet{1998A&A...340....1D}.
The mentioned effect is schematically illustrated in fig.~1 of \citet{2009ApJ...694..512M}.
It is probably also reflected on the corresponding distributions
of CC~SNe versus SNe~Ia in GD galaxies.

In addition, in GD hosts, there is a statistically significant
positive correlation for \textit{arm} SNe~II
between $d_2$ and $R_{\rm SN}/R_{\rm 25}$
(galactocentric distance of SN normalized to $R_{\rm 25}$ radius
of the galaxy).\footnote{{\footnotesize The $R_{25}$ is the SDSS $g$-band $25^{\rm th}$
magnitude isophotal semimajor axis of SN host galaxy.}}
The same correlation is found for SNe~Ibc,
with an even higher slope, but not significant
because of small number statistics.
The positions of both types of SNe beyond $R_{\rm SN}/R_{\rm 25}~\approx~0.45$
(roughly the mean corotation radius) are now typically outside the peaks of the arms
through the radius vector, while at smaller radii
the positions of SNe are typically inside the peaks of the arms.
Similar trends for star-forming regions are observed
in some GD galaxies \citep[e.g.][]{2013A&A...560A..59C}.

Adopting an average corotation radius of $0.45~R_{\rm 25}$,
the mean distances ($d_1$ inside and $1-d_1$ outside
the corotation radius, respectively)
of SNe~Ibc and II from leading edges of spiral arms
are $0.25 \pm 0.07$ and $0.44 \pm 0.03$, respectively. 
AD test shows that the difference
between the distributions of SNe~Ibc and II
relative to leading edges of spiral arms
is statistically significant ($P_{\rm AD} = 0.011$).
According to the dynamical simulations by \citet{2010MNRAS.409..396D},
the observed concentration sequence towards the leading edges of spiral arms indicates
a lifetime sequence \citep[see the top-left panel of fig.~4 in][]{2010MNRAS.409..396D}
for their progenitors (from youngest to oldest).
Hence, the greater the mean distance from the leading edge,
the longer is the progenitor's lifetime.

\begin{figure}
  \begin{center}$
  \begin{array}{c}
  \includegraphics[width=0.6\hsize]{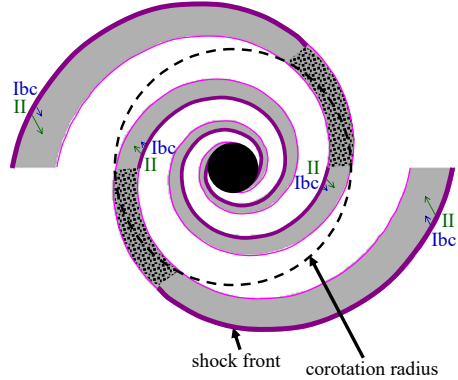}
  \end{array}$
  \end{center}
  \caption{The scheme of star formation distribution in a model of two armed GD galaxy
	         with the directions and relative sizes of drifts
           from birth places up to the explosion
					 for SNe~Ibc (blue arrow) and II (green arrow).
					 For better visualization, the directions of drifts are shown with
					 a significant radial component.}
  \label{GDchart}
\end{figure}

The scheme of star formation in a model of a GD galaxy with two spiral arms
with the directions and relative sizes of drifts
from birth places up to the explosion
for various SNe is given in the Fig.~\ref{GDchart}.
Inside the corotation radius (dashed circle), star
formation processes generally occur in a shock front
at the inner (leading) edges of spiral arms.
Since  the disc rotates faster than the spiral arms inside the corotation radius,
newborn stars near inner edges move towards
the outer edges of spiral arms. On the contrary, outside the corotation
radius, stars are caught up by the spiral arms, hence move from the outer
edges of the arms towards the inner edges of spiral arms.
In the corotation zone, there are no triggering mechanisms of star formation, such as
spiral shocks. The main mechanism of star formation in this region (dotted surface of arms) is
gravitation instability (as in NGD galaxies). Therefore, in this region, the distribution
of SNe inside the spiral arms should have the same behavior as in NGD galaxies.
Moreover, because of the absence of spiral shocks in this region,
star formation \citep[e.g.][]{1992ApJS...79...37E},
hence the number of CC SNe, should exhibit a drop.
Since more massive stars live shorter than less massive ones,
their explosion sites are, on average, closer to the leading edges of arms
where they born.
The observed significantly shorter distances
of SNe~Ibc from the leading edges of spiral arms show
that their progenitors are younger (more massive) than those of SNe~II.
This result is in agreement with the single-star progenitor scenario of SNe~Ibc.

To sum-up, the reported results show that the distribution
of SNe relative to spiral arms is a powerful tool
to constrain the lifetimes (masses) of their progenitors
and to better understand the star formation processes
in various types of spiral galaxies.

\acknowledgements This work was supported by the RA MES State Committee of Science,
in the frames of the research project number 15T--1C129.
This work was made possible in part by a research grant from the
Armenian National Science and Education Fund (ANSEF) based in New York, USA.



\end{document}